\documentclass[twocolumn,aps,prc,superscriptaddress,showpacs]{revtex4}
\usepackage{amsmath,bm}
\usepackage{graphicx}
\usepackage{dcolumn}
\usepackage{multirow}
\begin{document}
\draft
\title{Fluctuation of the download network}

\author{D. D. Han}
\affiliation{\footnotesize College of Information, East China
Normal University, Shanghai 200062, China}
\author{ J. G. Liu}
\affiliation{\footnotesize College of Information, East China
Normal University, Shanghai 200062, China}

\author{ Y. G. Ma}


\affiliation{\footnotesize Shanghai Institute of Applied Physics,
Chinese Academy of Sciences,  Shanghai 201800, China}


\date{\today}
\nopagebreak

\begin{abstract}

The scaling behavior of fluctuation for a download network which
we have investigated a few years ago based upon Zhang's
Encophysics web page has been presented. A power law scaling,
namely $\sigma \sim \langle f\rangle ^ \alpha $ exists between the
dispersion $\sigma$ and average flux $\langle f \rangle$ of the
download rates. The fluctuation exponent $\alpha$ is neither 1/2
nor 1 which was claimed as two universal fluctuation classes in
previous publication, instead it varies from 1/2 to 1  with the
time window in which the download data were accumulated. The
crossover behavior of  fluctuation exponents  can be qualitatively
understood by the external driving fluctuation model  for a
small-size system or a network traffic model which suggests
congestion as the origin.



\end{abstract}
\pacs{ 89.75.Hc, 89.75.Da, 89.40.Dd}

\maketitle

Many phenomenological and statistical analysis have been made for
the complex networks \cite{BA1,Dor}. In those researches, most
studies focused on the long-time behavior of a certain complex
network. In this sense, the feature of the network corresponds to
its static characteristic. However, time evolution of the network
topology is also very important. During its evolution, the network
nodes experience different traffic flux time by time and the
fluctuation is unavoidable. Actually, fluctuation is a universal
phenomenon which exists in many different fields, such as nuclear
fragmentation or hadron production \cite{Botet,Ma2,Gul}, which can
also be related to the critical behavior or self-organized
criticality. For instance, the dispersion ($\sigma$) of an order
parameter, such as the charge of the largest fragments in nuclear
fragmentation, shows a transition from $\sigma \propto \langle f
\rangle^{1/2}$ (the ordered phase) to $\sigma \propto \langle f
\rangle$ (the disordered phase) when the multifragmentation phase
transition takes place in hot nuclear system
 \cite{Botet,Ma2} (here
$\langle f \rangle$ is the average of the order parameter). For
network dynamics,  recently, Menezes and Barab\'asi investigated
the fluctuation in a number of real world networks, which includes
internet, river network, microchip, WWW and highway network
dynamics and presented a model to understand the origin of
fluctuation in traffic process \cite{Bara}. They found that the
fluctuation is dominantly driven by either internal or external
dynamics of the complex system \cite{Bara}. In their studies, they
found there is a power-law scaling for the dispersion and the
average flux, namely $\sigma \propto \langle f \rangle ^\alpha$,
and  there are two classes of universality for real systems. In
the Internet and the computer chip there is robust internal
dynamics which leads to the fluctuation exponent $\alpha$ = 1/2,
while highway and Web traffic are driven by external demand which
leads to the fluctuation exponent $\alpha$ = 1. Authors use a
stylized model of random walkers throughout network, they thought
what is probably one of the most important factors in the traffic
dynamics on networks is the limited capacity of nodes to handle
packets simultaneously, which leads to pack-pack interaction and
induce large fluctuations or even network congestion. However, a
recent study on scaling of fluctuation in internet traffic shows
that the fluctuation is different from 1/2 which was claimed in
the above papers. They developed a model where the arrival and
departure of "packets" follow exponential distribution, and the
processing capability of nodes is either unlimited or finite was
proposed by Duch and Arenas \cite{Duch}. This model presents a
wide variety of exponents between 1/2 and 1, revealing their
dependence on the few parameters considered, and questioning the
existence of universality classes. Hence it seems that the
universal classes of fluctuation scaling for network dynamics are
far from reaching consensus and therefore it is worthy to further
investigate what about the fluctuation behavior in other real
networks. Neverthless, so far there are few analysis on the
fluctuation behavior of other specific networks rather than the
networks which have been investigated in Ref.~\cite{Bara,Duch}. In
this work, we will investigate the network evolution and
fluctuation based on our previous study of the download network.

In our previous work in 2004 \cite{Han}, we reported, for the
first time, that the download frequency of the papers  in a web
page is also a scale-free network. Its  rank-ordered download
distribution can be described by the Zipf law \cite{Zipf_law,Ma1}
or Tsallis' non-extensive entropy \cite{Tsa}. The data set of the
download rates comes from a well constructed web page in the field
of economical physics (so-called Econophysics) by Zhang since 1998
\cite{Zhang_web}.
 Furthermore, the mechanism of network growth
was explained by the preferential attachment network model of
Barabasi and Albert. Since three years have passed after this
network analysis, it is of interesting to see how this network
evolves and how about the fluctuation of download rate.

\begin{figure}
\resizebox{19.2pc}{!}{\includegraphics{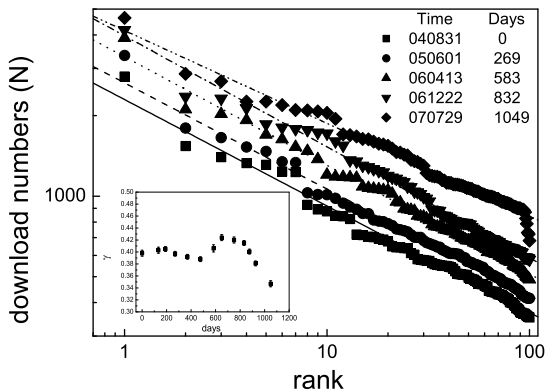}}
\vspace{-.8cm}\caption{\footnotesize The rank-ordered (Zipf-type)
plot for the download numbers of the papers in
http://www.unifr.ch/econophysics web page. The symbols are
illustrated in figure. See text for details. In the insert of left
bottom corner, it shows the evolution of Zipf exponent as a
function of the days starting from 2004/08/31. }
 \label{download}
\end{figure}

Firstly let us see some plots of  rank distributions of the
download numbers from the data on 2004/08/31 to 2007/07/29 which
is shown in Fig.~\ref{download}. Roughly, rank distribution are
almost linear in double logarithm plots and they can be described
by the Zipf law \cite{Zipf_law}
\begin{equation}
N \simeq rank^{-\gamma},
\end{equation}
where the $\gamma$ is the Zipf law exponent. Zipf's law or scale
free networks is different from the predictions of pure random
networks introduced by Erdos and Renyi \cite{Renyi}. Roughly, the
shapes of these distributions keep similar in all times. However,
quantitative analysis shows non-constant behavior of the evolution
of Zipf exponent (${\gamma}$) which is shown in the inset of the
Figure 1. Especially there is a bump during 2006, i.e. the
rank-ordered distributions tend to be steeper, which reflects
higher download frequency for higher rank papers. However, the
exponent decreases in 2007, i.e. more flatter distribution, which
is obviously seen in the Figure 1 (diamond points). In this case,
the web visitors prefer to download more papers listed in the web
page which are not only focused on those top downloaded papers.

\begin{figure}
\vspace{-0.5cm}
\resizebox{20.pc}{!}{\includegraphics{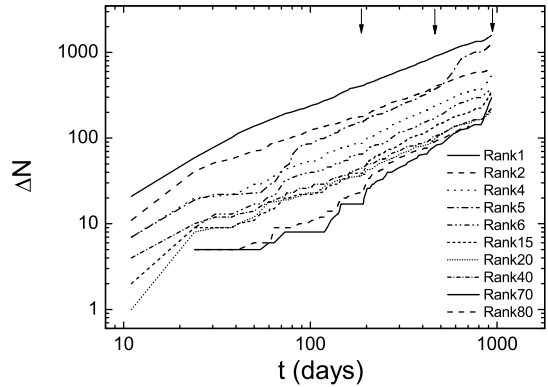}}
\vspace{-.8cm}\caption{\footnotesize Time evolution of download
numbers which is sorted by the different ranks of the papers. The
arrows in the upper axis illustrate three time points, namely
2005/03/13, 2005/12/26 and 2007/03/26 from left to right) which
will be used to investigate the time window effects afterwards. }
\label{Delta_N}
\end{figure}

To quantitatively see the increasing download numbers with time,
we make a plot in Fig.~\ref{Delta_N} for the rank-sorted download
numbers $\Delta N$ starting from the date 2004/08/31 (i.e. $\Delta
N$ = 0 on 2004/08/31) as a function of days which passed starting
from 2004/08/31. From the figure, all curves do not follow the
exact linear increasing. In other words, there exists fluctuation
for the download rates day by day.

Quantitative fluctuation of the download rates can be described by
the average download numbers per day, namely the average flux,
$\langle f \rangle = \frac{\Delta N}{t_{i+1}-t_i}$, where $t_i$ is
the time of the download day $i$ (i.e., the abscissa of
Fig.~\ref{Delta_N}) and $i$ from the starting date (2004/08/31) to
the ending date which will be illustrated later. For each
rank-ordered paper, these download rates change day by day,  from
which we can extract the average flux $\langle f\rangle$ and its
dispersion $\sigma$ (root of mean square of the download rate
distribution) for each paper.

\begin{figure}
\vspace{-.5cm} \resizebox{20.pc}{!}{\includegraphics{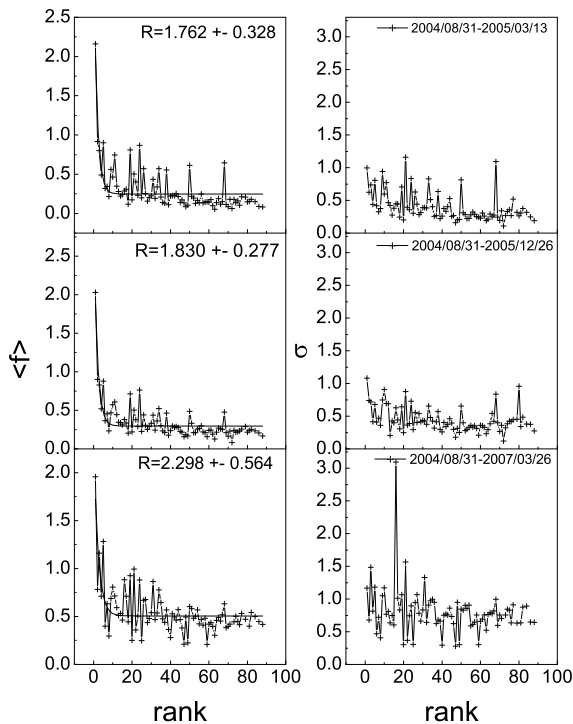}}
\vspace{-.8cm} \caption{\footnotesize The average flux $\langle f
\rangle$ (left column) or the dispersion $\sigma$ (left column) as
a function of the rank for the accumulated data during 2004/08/31
to 2005/03/13 (top row), from 2004/08/31 to 2005/12/26 (middle
row) and from 2004/08/31 to 2007/03/26 (bottom row). The line in
left panels represents the exponential fits using
Eq.\ref{eq-decay}.  } \label{rank-f}
\end{figure}

Fig.~\ref{rank-f} shows the relationship between the average
download rates $\langle f \rangle$ (left column) or the dispersion
$\sigma$ (left column) as a function of the rank for the
accumulated data during 2004/08/31 to 2005/03/13 (i.e. $\sim$ 6.5
months)(top row), from 2004/08/31 to 2005/12/26  (i.e. $\sim$ 31
months)(i.e. $\sim$ 16 months) (middle row)  and from 2004/08/31
to 2007/03/26  (i.e. $\sim$ 31 months) (bottom row). In left
columns, the download rates show a fast decay with the increasing
of  rank for those most downloaded papers and the keep fluctuation
for large rank values. It can be qualitatively understood that the
web visitors prefer to download the top rank-ordered papers when
he/she visits this page for the first time. This average
day-by-day download flux can be roughly described by the
exponential decay fits:
\begin{equation}
\langle f \rangle \propto exp(-\frac{rank}{R}),
\label{eq-decay}
\end{equation}
which is plotted in the figures and the half-lifetime decay
exponent $R$ is shown in the inset. $R$ is small and seems to
increase with the time period during which the data were
accumulated. Right columns depict the dispersion as a function of
the ranks which does not exhibit an obvious exponential decay as
$\langle f \rangle$ versus rank shows, rather than frequent
fluctuations.

\begin{figure}
\resizebox{20.pc}{!}{\includegraphics{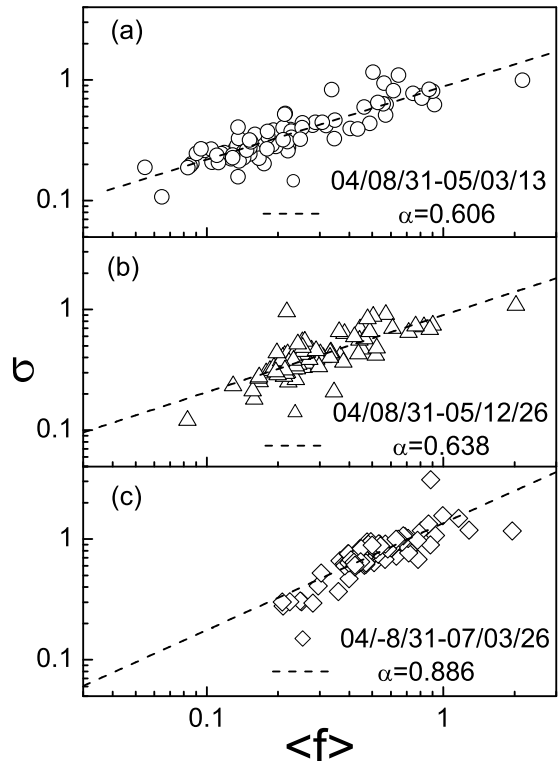}}
\vspace{-1.cm} \caption{\footnotesize $\sigma$ as a function of
$\langle f\rangle$  for all rank-ordered papers in three
accumulated time windows as illustrated in the figures.}
\label{scaling}
\end{figure}

Fig.~\ref{scaling} demonstrates the relationship of $\langle
f\rangle$ and $\sigma$ for all rank-ordered papers. To investigate
the possible effect of time window in which the download data were
accumulated, we use the data ensembles which correspond to the
period from 2004/08/31 to 2005/03/13  (a), from 2004/08/31 to
2005/12/26 (b) and from 2004/08/31 to 2007/03/26 (c),
respectively.  From each double logarithm plot, all points
basically show linear increases between the average flux and its
dispersion. In this context, we fit the data points using the
power law:
\begin{equation}
\sigma \propto \langle f\rangle ^ \alpha
\end{equation}
to extract the scaling parameter $\alpha$ which are shown in the
inset of each panel. There are two points which  we can learn from
the figure: (1) the scaling parameter $\alpha$ is neither 1/2 nor
1. In the work of Menezes and Barab\'asi, they thought there are
two universal fluctuation classes: $\alpha \simeq 1/2$ or 1
systems. The typical example of the former is the Internet
network, which was claimed to be dominantly driven by the internal
dynamics. And the typical example of the latter is the WWW URL
links, which was claimed to be dominantly driven by the external
dynamics. The exponents of our download network are between 1/2
and 1. (2) The scaling exponents seem to depend on the time
windows during which the data samples are collected. In the other
words,  the longer the time windows, the larger the fluctuation
exponent. Hence, in this viewpoint, we cannot exclude the
possibility that fluctuation exponent could reach to 1 if we take
very long time windows from the present work.

Qualitatively, our network fluctuation could be explained by the
interplay of different fluctuation types, namely internal
fluctuation and external fluctuation, which was proposed by
Menezes and Barab\'asi \cite{Bara}. In their model, they consider
the random diffusion of $W$ walkers (here they represent the
visitors who download the papers) on the network, such that each
walker that reaches a node $i$ (here it represents the
rank-ordered paper) departs in the next time step along one of
other nodes. Originally each walker is placed on the network at a
randomly chosen location and removed after it performs $M$ steps,
mimicking in a highly simplified fashion a human browser surfing
the Web for information \cite{Bara}. In this way, finally the
relationship between the average flux and the fluctuations follow
a fluctuation scaling with $\alpha$ = 1/2, corresponding to
internal fluctuation driven behavior. However, in real systems the
fluctuation on a given node is determined not only by the system's
internal dynamics, but also by changes in the external condition.
To incorporate externally induced fluctuation, Menezes and
Barab\'asi allow $W$ (the number of walkers in the web page), to
vary from one day to the other. This is of course true, especially
in case that peoples visit un-congested web page, such as the
Encophysics web page. Assuming that the day to day variations of
$W(t)$ define a dynamic variable chosen from an uniform
distribution in the interval [$W$-$\Delta W$, $W+ \Delta W$], for
$\Delta W$ = 0 one recovers $\alpha$ = 1/2. However, when $\Delta
W$ exceeds a certain threshold, in both models the dynamical
exponent changes to $\alpha$ = 1 \cite{Bara}. In this case, the
external fluctuation can overshadow the internal fluctuation so
that $\alpha$ = 1. However, our network fluctuation behavior is
not exactly the above extreme cases, instead it seems to be
located in the transition region between the two extreme cases,
namely the internal fluctuation type ($\alpha$ = 1/2) and external
fluctuation type ( $\alpha$ = 1). This could be explained by a
smaller $\Delta W$ which does not exceed the certain threshold
corresponding to a transition condition from $\alpha$ = 1/2 to 1
in our download network. We think this is reasonable since the
Encophysics web page is not a popular web page, such as Yahoo or
Sina web pages, instead it is a small-circle scientific web page
and no many people browse it often. In this case, $\Delta W$ could
be small due to a few people browse this web page day-by-day so
that the day variations $\Delta W$ cannot exceed the certain
threshold. Actually, there exists a wide transition region where
$\alpha$ is between 1/2 to 1 for a  finite network system in
Menezes and Barab\'asi's model. Therefore, our interpretation for
the origin of the download network fluctuation is not contradicted
to their model. Using the above scenario, we can qualitatively
learn what the fluctuation originates from for our download
network.

However, the above scenario  which assumes any external driving
force is not unique to interpret the observed crossover
fluctuation exponents between 1/2 and 1. A simple traffic model in
complex networks that suggests congestion as the origin of the
increase of $\alpha$ and captures the essential parameters
governing the dynamical process  \cite{Duch} is also possible to
explain the download network fluctuation. In that model, traffic
process in a complex network of $N$ nodes as $N$ queue systems,
and a random walk simulation for the movement of packets on the
network. The arrival process of packets to the network is
controlled by a Poisson distribution with parameter $\lambda$,
each packet enters the network at a random selected node. Once the
packet arrives to the node enters a queue. The delivery of the
packets in the queue is controlled by an exponential distribution
of service times with parameter $\mu$. In that model, the packets
will perform $S$ random steps in the network before disappearing.
This dynamics is performed in continuous time, assuming that the
time expended by packets traveling through a link is negligible.
The model can finally account for different scaling exponents
$\alpha$ depending on the parameters $\lambda$, $\mu$, $S$, and
the time period $P$. Especially, we are interesting to see that
$\alpha$ is a function of the time window length $P$ in which the
average was taken, which changes from 1/2 to 1: $\alpha$ increases
with the the time window length in the transition region. In the
present study, even though our time window means the whole
statistical one in which the download data were accumulated, which
is different from the above mentioned time window in which the
average were taken in the above model, the effect could be
analogous: the larger accumulated time windows can be somewhat
equivalent to the larger time window length $P$ in which the
average were taken. The same trend which $\alpha$ increases with
time was observed. Actually, the fluctuation exponents between 1/2
and 1 have been observed in the stock market transaction and other
human dynamics such as emails from a particular company and data
on the printing activity etc, and their exponent shows the
dependences on the time window size. A detailed review can be
found in Ref.~\cite{rev}.

In summary, we investigated the evolution of the download network
for the rank-ordered papers which were listed in Zhang's
Encophysics web page. In recent three years, the download
distribution shows the change of the exponents even though the
rank-ordered distribution still keeps scale-free feature,
reflecting the change of traffic on nodes which represent the
given downloaded papers. Further, we give quantitative analysis
for the average download rates $\langle f \rangle$ per day, which
show day-by-day fluctuation. The average flux shows a fast
exponential decay as a function of the rank, while the dispersion
does not show an obvious dependence of the rank.  Interestingly,
the dispersion of the download rate distributions shows a
power-law scaling behavior with its average flux, namely $\sigma
\propto \langle f \rangle ^{\alpha}$. In different time windows
ranging from about 6.5 months to 31 months in which the download
distributions are accumulated, the scaling parameter $\alpha$
changes with the time windows, namely from 0.60 to 0.89. The
origins are qualitatively interpreted by two models. Future work
on quantitative model simulation and a possible  $\Delta$-scaling
 of network fluctuation is in progress.

This work was partially supported by Shanghai Development
Foundation for Science and Technology under Grant Numbers
06JC14082 and 05XD14021.

 {}
\end{document}